\documentclass[aps,prl,reprint]{revtex4-1}
\usepackage{blindtext}
\usepackage{graphicx}

\begin{document}
\def\bigint{{\displaystyle\int}}
\def\simlt{\stackrel{<}{{}_\sim}}
\def\simgt{\stackrel{\rangle}{{}_\sim}}

\title{
Solving the Greenberger-Horne-Zeilinger paradox: an explicitly local and realistic model of hidden variables for the GHZ quantum state 
}

\author{David H. Oaknin}
\affiliation{Rafael Ltd, IL-31021 Haifa, Israel \\
e-mail: d1306av@gmail.com \\
}

\begin{abstract}
The Greenberger-Horne-Zeilinger~(GHZ) version of the Einstein-Podolsky-Rosen~(EPR)  paradox  is widely regarded as a conclusive logical argument that rules out the possibility of describing quantum phenomena within the framework of a local and realistic model of hidden variables in which the observers are free to choose their own experimental settings. In this paper we show, however, that the GHZ argument implicitly relies on an additional crucial assumption, which is not required by fundamental physical principles and had gone unnoticed. Namely, we note that the argument implicitly assumes the existence of an absolute angular frame of reference with respect to which the polarization properties of the hypothetical hidden configurations of the entangled particles as well as the orientation of the measurement apparatus that test the system can be defined. We further note that such an absolute frame of reference would not exist if the hidden configurations of the entangled particles spontaneously break the gauge rotational symmetry. Indeed, by skipping this unnecessary additional assumption we are able to build an explicitly local and realistic model of hidden variables for the GHZ state, which complies with the 'free-will' hypothesis and reproduces the quantum mechanical predictions, and thus completes the description of the system in the EPR sense. 
%Our model directly links the {\it elements of reality} to the {\it weak values} of the quantum observables that describe them. 
\end{abstract}

\maketitle

{\bf 1.}  The inability to accommodate the seemingly trivial notions of locality and physical realism within the current interpretation of the quantum mechanical wavefunction is at the core of a long lasting debate about the foundations of the quantum theory and the role played by measurements, whose origins go back to the formulation of the renowned Einstein-Podolsky-Rosen~(EPR) paradox seventy years ago \cite{EPR}. Solving these key issues would require to develop a description of quantum phenomena in terms of a local and realist model of hidden variables. Nonetheless, according to the current wisdom, such a description is not possible in so far as we insist on keeping the notion, also seemingly trivial, that the observers' choice of their measurement settings is not constrained by the actual hidden configuration of the observed system (free-will). Indeed, several fundamental theorems state that generic models of hidden variables that share certain intuitive features cannot fully reproduce the predictions of quantum mechanics \cite{Bell, K-S, CHSH, CH, GHZ, Cabello1, Leggett, Colbeck, Fine}. Moreover, carefully designed experimental tests have consistently confirmed the predictions of the quantum theory and, thus, ruled out these generic models of hidden variables \cite{Aspect, Tittel, Weihs, Matsukevich, Rowe, Zeilinger, Branciard, Cabello2, Cabello3, Hansen}. 

The best known among these theorems is Bell's theorem \cite{Bell,CHSH,CH,Fine}, which proves that such generic models of hidden variables cannot reproduce the statistical correlations predicted by quantum mechanics for the outcomes of long sequences of strong polarization measurements performed along certain relative directions on two entangled particles, if the measurements are fully anti-correlated when performed along {\it parallel} directions.  

The  Greenberger-Horne-Zeilinger version of Bell's theorem \cite{GHZ} is an even more conclusive proof of the limitations of these generic local and realistic models of hidden variables, since it proves that such models cannot reproduce even single outcomes of strong spin polarization measurements performed along certain relative directions on three or more entangled particles prepared in the so-called $\mbox{GHZ}$ state, 
\begin{equation}
|\mbox{GHZ} \rangle \equiv \frac{|\uparrow \ \uparrow \ ..... \ \uparrow \ \rangle  + |\downarrow \ \downarrow \ ..... \ \downarrow \ \rangle}{\sqrt{2}}. 
\end{equation}

However, in a recent paper \cite{oaknin3} we have shown that Bell's theorem crucially relies on an implicit feature of the considered generic class of models of hidden variables, beyond the aforementioned features of locality, physical realism and 'free-will'. This additional feature, unlike the other three explicit features, is not required by fundamental physical principles. Namely, we noted that Bell's theorem implicitly assumes that there exists an absolute angular frame of reference with respect to which we can define the polarization properties of the hypothetical hidden configurations of the pair of entangled particles as well as the orientation of the meassurement apparatus that test them. Indeed, we showed that this implicit assumption is not correct if these hidden configurations spontaneously break the gauge rotational symmetry along an otherwise arbitrary direction. Following this insight we actually built an explicitly local and realistic model of hidden variables that complies with the hypothesis of 'free-will' and fully reproduces the predictions of quantum mechanics for Bell's states and, thus, it completes the description of the quantum polarization singlet in the sense advocated by Einstein, Podolsky and Rosen in \cite{EPR}. 

In this paper we develop these ideas to build an explicitly local and realistic model of hidden variables for the GHZ state of three entangled particles. The paper is organized as follows. In Section 2 we review the argument put forward by Greenberger, Horne and Zeilinger against the impossibility to describe the quantum mechanical predictions for the GHZ state within the framework of a local and realistic model of hidden variables. In Section 3 we introduce a simple explicit model of hidden variables that overcomes this argument. In Section 4 we extend this model and discuss it in detail. Our conclussions are summarized in section 5. 
\\

%In particular, we notice that the polarization properties of the hidden configurations can only be properly defined with respect to %the orientation of the measurement apparatus, each of which define an angular frame of reference. The polarization properties of %these hidden configurations may actually take different values when described with respect to different frames of reference, and %yet the model be counter-factual and non-contextual. Indeed, there does not exists any constraint that prevent the actual values %of the polarization components of, say, particle A to depend on the orientation of the reference direction with respect to which we %describe them. Moreover, only the polarization component along the reference direction must take a binary value, either $+1$ or %$-1$, and obey standard algebraic relationships. Other components are not necessarily contrained to comply with these %relationships and could, in principle, take even complex values.

{\bf 2.} The Greenberger-Horne-Zeilinger spin polarization state of three entangled particles is described by the quantum wavefunction:
\begin{eqnarray}
\label{GHZ}
\nonumber
|\Pi\rangle=\frac{1}{\sqrt{2}}\left(|\uparrow\rangle^{(A)} \otimes |\uparrow\rangle^{(B)} \otimes |\uparrow\rangle^{(C)} \ \  + \hspace{0.3in} \right. \\
\left.  + \ \  e^{i \Phi} \ |\downarrow\rangle^{(A)} \otimes |\downarrow\rangle^{(B)} \otimes |\downarrow\rangle^{(C)}\right),
\end{eqnarray}
where $\{|\uparrow\rangle, |\downarrow\rangle\}$ denotes a basis of single particle spin polarization eigenstates along its locally defined Z-axis. In this state all three outcomes in each single event of a long sequence of strong spin polarization measurements performed on each of the three particles along their corresponding Z-axes must be consistently equal, either 
\begin{eqnarray*}
S^{(A)}_Z(n)=S^{(B)}_Z(n)=S^{(C)}_Z(n)= +1,
\end{eqnarray*}
or 
\begin{eqnarray*}
S^{(A)}_Z(n)=S^{(B)}_Z(n)=S^{(C)}_Z(n)= -1,
\end{eqnarray*} 
for all $n \in \{1,...,N\}$, with each one of the two possibilities happening with a probability of 1/2.

In fact, in the GHZ state (\ref{GHZ}) the expected average values of long sequences of strong spin polarization measurements performed along any arbitrary directions $\Omega^{(A)}_{\alpha}$, $\Omega^{(B)}_{\beta}$, $\Omega^{(C)}_{\gamma}$ in the XY-planes orthogonal to the local Z-axes are equal to zero:

\begin{equation}
\langle S^{(A)}_{\Omega_{\alpha}}(n) \rangle_{n \in {\bf N}} = \langle S^{(B)}_{\Omega_{\beta}}(n) \rangle_{n \in {\bf N}} = \langle S^{(C)}_{\Omega_{\gamma}}(n) \rangle_{n \in {\bf N}} = 0,
\end{equation}
as well as their two-particles correlations: 
\begin{equation}
\begin{array}{c}
\langle S^{(A)}_{\Omega_{\alpha}}(n) \cdot S^{(B)}_{\Omega_{\beta}}(n) \rangle_{n \in {\bf N}} = \langle S^{(B)}_{\Omega_{\beta}}(n) \cdot S^{(C)}_{\Omega_{\gamma}}(n) \rangle_{n \in {\bf N}} = \\ \hspace{1.3in} = \langle S^{(C)}_{\Omega_{\gamma}}(n) \cdot S^{(A)}_{\Omega_{\alpha}}(n) \rangle_{n \in {\bf N}} = 0.
\end{array}
\end{equation}
Notwithstanding, the three-particles correlation is non-zero, in general, and given by: 

\begin{eqnarray}
\label{the_law}
\nonumber
\langle S^{(A)}_{\Omega_{\alpha}}(n) \cdot S^{(B)}_{\Omega_{\beta}}(n) \cdot S^{(C)}_{\Omega_{\gamma}}(n) \rangle_{n \in {\bf N}} = \hspace{0.7in} \\
\nonumber
= \cos\left(\Delta_{\Omega^{(A)}_{\alpha}} + \Delta_{\Omega^{(B)}_{\beta}} + \Delta_{\Omega^{(C)}_{\gamma}} + \Phi\right), 
\end{eqnarray}
where $\Delta_{\Omega^{(A)}_{\alpha}}$, $\Delta_{\Omega^{(B)}_{\beta}}$ and $\Delta_{\Omega^{(C)}_{\gamma}}$  describe the relative orientations of each one of the measurement apparatus with respect to some implicit reference direction, which we define as X-axis. 

In particular,for $\Phi=0$  the following four relationships follow:

\begin{equation}
\label{4EPR}
\begin{array}{cccccccccccccc}
S_X^{(A)}(n) & \cdot & S_X^{(B)}(n) & \cdot & S_{X}^{(C)}(n)  & = +1, &n=1,....,N\\
S_X^{(A)}(m) & \cdot & S_Y^{(B)}(m) & \cdot & S_{Y}^{(C)}(m) & = -1, &m=1,....,M\\
S_Y^{(A)}(k) & \cdot & S_X^{(B)}(k) & \cdot & S_{Y}^{(C)}(k) & = -1, &k=1,....,K\\
S_Y^{(A)}(l) & \cdot & S_Y^{(B)}(l) & \cdot & S_{X}^{(C)}(l) & = -1, &l=1,....,L,
\end{array}
\end{equation}
for any four sequences of strong measurements performed along directions $(X,X,X)$, $(X,Y,Y)$, $(Y,X,Y)$ and $(Y,Y,X)$.

These four relationships (\ref{4EPR}) lie at the core of the Greenberger-Horne-Zeilinger paradox \cite{GHZ}. On one hand, these relationships imply that we can gain certainty about the polarization properties of any of these three particles without in any sense disturbing them. Thus, according to the notion introduced by Einstein, Podolsky and Rosen \cite{EPR}, these polarization properties are {\it elements of reality} whose values must be set at the time when the three entangled particles are produced. On the other hand, this notion seems to be inconsistent: by multiplying the last three equations in (\ref{4EPR}) and assuming that all polarizations components must take values either $+1$ or $-1$, we would obtain that 

\begin{equation}
\begin{array}{cccccccccccccccccccccccccc}
S_X^{(A)}(n) & \cdot & S_X^{(B)}(n) & \cdot & S_{X}^{(C)}(n)  & = -1, &n=1,....,N
\end{array}
\end{equation}   
which is in contradiction with the first one. 

This argument is widely considered as the most clear-cut evidence against the possibility of giving the wavefunction (\ref{GHZ}) an statistical interpretation within the framework of a local and realistic model of hidden configurations, in which the observers are free to choose the setting of their measurements.
\\

{\bf 3.} The above argument crucially relies on the implicitly assumed existence of an absolute angular frame of reference, with respect to which the polarization properties of the hidden configurations of the triplet of entangled particles, as well as the orientation of the three measurement apparatus that test them, can be defined. In such an absolute frame of reference all the polarization components of each possible hidden configuration must take a binary value, either $+1$ or $-1$,  and relationships (\ref{4EPR}) immediately follow. However, as we already did notice in \cite{oaknin3}, the existence of such an absolute angular frame of reference is not required by fundamental physical principles. Indeed, the polarization properties of the hidden configurations can only be properly defined with respect to the reference directions set by the orientation of the measurement apparatus. Furthermore, the orientation of the measurement apparatus can only be defined with respect to each other, while their global orientation is actually an spurious gauge degree of freedom.  

That is, the orientation of the three measurement apparatus, whose correlation is given by 

\begin{equation}
\label{the_law}
\langle S^{(A)}_{\Omega_{\alpha}}(n) \cdot S^{(B)}_{\Omega_{\beta}}(n) \cdot S^{(C)}_{\Omega_{\gamma}}(n) \rangle_{n \in {\bf N}} = \cos\left(\Delta + \Phi\right),
\end{equation}
is described by a single physical degree of freedom, namely, the angle $\Delta$ measured with respect to a set of X-axes locally defined at the sites of each one of the three particles and for which the correlation is given by:

\begin{equation}
\label{the_frame}
\langle S^{(A)}_X(n) \cdot S^{(B)}_X(n) \cdot S^{(C)}_X(n) \rangle_{n \in {\bf N}} = \cos(\Phi),
\end{equation}
and, in particular for $\Phi=0$ by
\begin{equation}
\label{the_frame0}
\langle S^{(A)}_X(n) \cdot S^{(B)}_X(n) \cdot S^{(C)}_X(n) \rangle_{n \in {\bf N}} = +1.
\end{equation}
Actually, condition (\ref{the_frame}) defines the notion of {\it parallel} directions at the sites of each one of the three entangled particles. All sets of axes for which this condition is fulfilled are gauge equivalent and, hence, physically undistinguishable through measurements performed on a triplet of entangled particles prepared in the GHZ state. 

Moreover, the polarization properties of the hidden configurations of the triplet of entangled particles can be properly defined only with respect to the local reference directions set by the three measurement apparatus. That is, the actual value $s^{(A)}_{\Omega}(\Omega_{\alpha}, \omega)$ of the polarization component of, say, particle A along some direction $\Omega$ may be, in general, a function  of the reference direction $\Omega_{\alpha}$ set by the measurement apparatus of observer A (and, of course, also of the coordinate $\omega \in {\cal S}$ that labels the hidden configuration in which the system of three entangled particles occurs). This dependence does not conflict with the principles of locality and realism, which only demand that the value of the polarization components of particle A cannot depend on the orientations of the reference directions $\Omega_{\beta}$, $\Omega_{\gamma}$  along which observers B and C choose to test their particles.
Therefore, we must not restrict our models within the constraint that all polarization components of either one of the particles must take a binary value, either $+1$ or $-1$: only the polarization component of each one of the particles along the reference direction set by the orientation of the corresponding measurement apparatus must take a binary value. That is, on all possible hidden configurations of the triplet we must have:

\begin{eqnarray}
\label{realism_New}
\nonumber
s^{(A)}_{\Omega_{\alpha}}(\Omega_{\alpha}, \omega) = \pm 1, \ \
\nonumber
s^{(B)}_{\Omega_{\beta}}(\Omega_{\beta}, \omega) = \pm 1, \ \
\nonumber
s^{(C)}_{\Omega_{\gamma}}(\Omega_{\gamma}, \omega) = \pm 1,
\end{eqnarray}
\begin{equation}
\label{realism}
\hspace{1.0in}
\end{equation}
but the polarization components along any other directions must not necessarily take either one of these two values. Indeed, the only experimental access that we can have to the spin polarization components along these other directions is through weak measurements, whose outcome can have absolute values larger and smaller than one and may even be complex \cite{Jozsa}.  

Therefore, it is crucial to realize that in order to obtain a meaningful description of the system we must be careful to compare magnitudes defined with respect to the same reference directions. For example, we can state that with respect to a set of {\it parallel} reference directions $X^{(A)}$, $X^{(B)}$ and $X^{(C)}$ defined by condition (\ref{the_frame0}), the polarization components of the particles along the orthogonal directions $Y^{(A)}$, $Y^{(B)}$, $Y^{(C)}$ take values either $+i$ or $-i$, according to the relationship: 

\begin{equation}
\label{My_statement}
\begin{array}{cccccccc}
s^{(A)}_Y(X, \omega) & = & i & s^{(A)}_X(X, \omega), \\ 
s^{(B)}_Y(X, \omega) & = & i & s^{(B)}_X(X, \omega), \\
s^{(C)}_Y(X, \omega) & = & i & s^{(C)}_X(X, \omega),
\end{array}
\end{equation} 
with $s^{(A)}_X(X, \omega) = \pm 1$, $s^{(B)}_X(X, \omega) = \pm 1$ and $s^{(C)}_X(X, \omega) =\pm 1$. Then, the four constraints (\ref{4EPR}) become trivially identical,

\begin{equation}
\label{4EPRB}
\begin{array}{cccccccccccccc}
\ \ s_X^{(A)}(X,\omega) & \cdot & \ \ s_X^{(B)}(X,\omega) & \cdot & \ \ s_{X}^{(C)}(X,\omega)  & = +1, \\
\ \ s_X^{(A)}(X,\omega) & \cdot & i \ s_X^{(B)}(X,\omega) & \cdot & i \ s_{X}^{(C)}(X,\omega) & = -1, \\
i \ s_X^{(A)}(X, \omega) & \cdot & \ \ s_X^{(B)}(X,\omega) & \cdot & i \ s_{X}^{(C)}(X,\omega) & = -1, \\
i \ s_X^{(A)}(X, \omega) & \cdot & i \ s_X^{(B)}(X,\omega) & \cdot & \ \ s_{X}^{(C)}(X,\omega) & = -1. 
\end{array}
\end{equation}
In other words, the argument put forward by Greenberger, Horne and Zeilinger as a proof of the impossibility to reproduce the predictions of quantum mechanics for the GHZ state within the framework of a local and realistic statistical model of hidden variables can be overcome by realizing that there does not necessarily exists an absolute frame of reference with respect to which the polarization properties of the entangled particles can be defined and, in consequence, allowing their actual values depend on the reference direction with respect to which they are described. 
%The polarization components along any other direction can be defined as linear combinations of these.
\\

{\bf 4.}  In this section we build and discuss in detail an explicitly local and realistic statistical model of hidden variables for the GHZ state of three entangled particles. The model complies with the 'free-will' assumption and reproduces the quantum mechanical predictions for the average values and correlations of long sequences of strong spin polarization measurements performed along any three arbitrary directions. 

Our statistical model consists of a continuous set ${\cal S}$ of possible hidden configurations labelled by a coordinate $\omega \in {\cal S}$, with a well-defined probability density ${\cal G}(\omega)$ for each one of these configurations to occur and a locally defined binary response function $S^{(A,B,C)}_{\Omega_{\alpha,\beta,\gamma}}(\omega)$ that specifies the outcome that each one of these hidden configurations would produce in each one of the three measurement apparatus as a function of their orientations.

%As we have noticed above it is physically meaningless to describe the measurement setting chosen by the observers by specifying %independently the angular orientations of each one of the three measurement apparatus: only their relative orientation $\Delta$ with %respect to a different setting that serves as reference (\ref{the_frame}) can be physically defined. Moreover, we must describe %the polarization properties of each particle in each possible hidden configuration with respect to the orientation of the %corresponding measurement apparatus.  A crucial point in our model will be the transformation laws that relate the polarization %properties of the hidden configurations with respect to different reference directions.
%Indeed, when the hidden configurations of the triplet of particles spontaneously break the rotational symmetry it does not make %sense either to define within the same framework the relative angle between the orientations of measurement apparatus B and C.  

In particular, the phase space of our model consists of infinitely many possible hidden configurations uniformly distributed over a unit torus ${\cal S}_1 \times {\cal C}_1$. An observer that strongly measures the polarization of, say, particle A along an arbitrary direction $\Omega^{(A)}_{\alpha}$ orthogonal to its locally defined Z-axis fixes a reference frame of angular coordinates $(\omega_A, \eta) \in [-\pi, \pi) \times [-\pi, \pi)$ over this torus. 

We assume that the probability density distribution of each one of the hidden configurations to occur is given by

\begin{equation}
\label{density_states}
G(\omega_A, \eta) =  \frac{1}{2\pi} \cdot g(\omega_A), \hspace{0.15in} \mbox{with} \hspace{0.1in} g(x) = \frac{1}{4} \left|\sin\left(x\right)\right|
\end{equation}
and the outcome of a strong measurement of the polarization component of particle A along the chosen reference direction is given by:

\begin{equation}
S^{(A)}_{\Omega_{\alpha}}(\omega_A,\eta) = s^{(A)}_{\Omega_{\alpha}}(\Omega_{\alpha},\omega_A,\eta),
\end{equation}
with:
 
\begin{eqnarray}
s^{(A)}_{\Omega_{\alpha}}(\Omega_{\alpha},\omega_A,\eta) = \left\{
\begin{array}{cccc}
\hspace{-0.05in}&\hspace{-0.07in}S(\omega_A), & \mbox{if} \hspace{0.1in} \eta \in (0, +\pi], \\
\hspace{-0.05in}-&\hspace{-0.07in}S(\omega_A), & \mbox{if} \hspace{0.1in} \eta \in (-\pi, 0],
\end{array}
\right.
\end{eqnarray}
and

\begin{eqnarray}
S(\omega) = \left\{
\begin{array}{ccccc}
+1, & \mbox{if} & \omega \in & (0, +\pi], \\
-1, & \mbox{if} & \omega \in & (-\pi, 0].
\end{array}
\right.
\end{eqnarray}
\

The observer of particle B measures its polarization along some other arbitrary direction $\Omega^{(B)}_{\beta}$ orthogonal to its local Z-axis, which sets a different reference direction with its own frame of angular coordinates $(\omega_B, \eta) \in [-\pi, +\pi) \times [-\pi, \pi)$ over the torus ${\cal S}_1 \times {\cal C}_1$. By symmetry considerations we demand that the outcome of this measurement is described by the same response function:

\begin{eqnarray*}
S^{(B)}_{\Omega_{\beta}}(\omega_B,\eta) = s^{(B)}_{\Omega_{\beta}}(\Omega_{\beta},\omega_B,\eta) = \hspace{0.6in} \\
 = \left\{
\begin{array}{cccc}
\hspace{-0.05in}&\hspace{-0.07in}S(\omega_B), & \mbox{if} \hspace{0.1in} \eta \in (0, +\pi], \\
\hspace{-0.05in}-&\hspace{-0.07in}S(\omega_B), & \mbox{if} \hspace{0.1in} \eta \in (-\pi, 0].
\end{array}
\right.
\end{eqnarray*}
The two sets of angular coordinates $\omega_A$ and $\omega_B$ are related by the relationship:

\begin{equation}
\label{parallelism}
\begin{array}{ccccccccccccc}
\omega_B & = & & &\omega'_A, & \hspace{0.15in} \mbox{if} & \hspace{0.1in} & \eta & \in & ( & 0 &, \pi], \\
\omega_B & = &\pi & + &\omega'_A, & \hspace{0.15in} \mbox{if} & \hspace{0.1in} & \eta & \in & ( & -\pi &, 0],
\end{array} 
\end{equation}
where

\begin{equation}
\omega'_A = L(\omega_A; \Delta),
\end{equation}
and

\begin{itemize}
\item If  $\Delta \in [0, \pi)$, 
\begin{eqnarray}
\label{Oaknin_transformation}
\hspace{-0.15in}
L(\omega; \Delta) =  
\left\{
\begin{array}{c}
\hspace{-0.2in} q(\omega) \cdot \mbox{acos}\left(-\cos(\Delta) - \cos(\omega) - 1 \right), \\ \hspace{0.88in} \mbox{if}  \hspace{0.1in} -\pi \hspace{0.16in} \le  \omega < \Delta-\pi, \\
\hspace{-0.2in} q(\omega) \cdot \mbox{acos}\left(+\cos(\Delta) + \cos(\omega) - 1 \right), \\ \hspace{0.685in} \mbox{if}  \hspace{0.05in} \Delta-\pi \hspace{0.08in} \le \omega < \hspace{0.105in} 0, \\
\hspace{-0.2in} q(\omega) \cdot \mbox{acos}\left(+\cos(\Delta) - \cos(\omega) + 1 \right), \\ \hspace{0.69in} \mbox{if}  \hspace{0.25in} 0 \hspace{0.18in} \le \omega < \ \Delta, \\
\hspace{-0.2in} q(\omega) \cdot \mbox{acos}\left(-\cos(\Delta) + \cos(\omega) + 1 \right), \\ \hspace{0.72in} \mbox{if}  \hspace{0.21in} \Delta  \hspace{0.19in} \le  \omega  < +\pi, \\
\end{array}
\right.
\end{eqnarray}
\item If  $\Delta \in [-\pi, 0)$, 
\begin{eqnarray}
\label{Oaknin_transformation_Inv}
\hspace{-0.15in}
L(\omega; \Delta) =  
\left\{
\begin{array}{c}
q(\omega) \cdot \mbox{acos}\left(-\cos(\Delta) + \cos(\omega) + 1 \right), \\ \hspace{0.650in} \mbox{if}   \hspace{0.11in} -\pi \hspace{0.15in} \le \omega < \Delta, \\
q(\omega) \cdot \mbox{acos}\left(+\cos(\Delta) - \cos(\omega) + 1 \right), \\ \hspace{0.66in} \mbox{if}   \hspace{0.30in} \Delta \hspace{0.1in} \le \omega < \hspace{0.05in} 0, \\
q(\omega) \cdot \mbox{acos}\left(+\cos(\Delta) + \cos(\omega) - 1 \right), \\ \hspace{0.92in} \mbox{if}  \hspace{0.23in} 0 \hspace{0.22in} \le \omega < \Delta +\pi, \\
q(\omega) \cdot \mbox{acos}\left(-\cos(\Delta) - \cos(\omega) - 1 \right), \\ \hspace{0.75in} \mbox{if}  \hspace{0.16in} \Delta +\pi \hspace{0.00in} \le \omega < +\pi, \\
\end{array}
\right.
\end{eqnarray}
\end{itemize}
with 
\begin{eqnarray*}
q(\omega) = \mbox{sign}((\omega - \Delta) \mbox{mod} ([-\pi, \pi))),
\end{eqnarray*}
and the function $y=\mbox{acos}(x)$ is defined in its main branch, such that $y \in [0, \pi]$ while $x \in [-1, +1]$. The parameter $\Delta$ in this transformation law denotes now the relative orientation of the measurement setting $\Delta + \Phi$, as defined in (\ref{the_law},\ref{the_frame}).
\\

It is straightforward to check that the probability density (\ref{density_states}) remains invariant under the coordinates transformation (\ref{Oaknin_transformation},\ref{Oaknin_transformation_Inv}), since 
\begin{eqnarray}
\nonumber
|\mbox{d}\omega' g(\omega')| = \frac{1}{4} |\mbox{d}\omega' \ \sin(\omega')| = \frac{1}{4} |\mbox{d}(\cos(\omega'))| = \\
= \frac{1}{4} |\mbox{d}(\cos(\omega))| = \frac{1}{4} |\mbox{d}\omega \ \sin(\omega)| = |\mbox{d}\omega \  g(\omega)|,
\end{eqnarray}
 and 
\begin{equation}
g(\pi+\omega) = \frac{1}{4} |\sin(\omega+\pi)| = \frac{1}{4} |\sin(\omega)| = g(\omega).
\end{equation}
In fact, these equalities state in precise terms that the probability of each hidden configuration to occur does not depend on the orientation of the reference direction chosen by the observers to describe their particles or, in other words, that our model complies with the requirements of 'free-will'.

We can now define a partition of the torus into four disjoint regions, ${\cal S}_1 \times {\cal C}_1 = {\cal I}_{++} \ \bigcup \ {\cal I}_{+-}  \ \bigcup \ {\cal I}_{-+}  \ \bigcup \ {\cal I}_{--}$, as follows:

\begin{eqnarray*}
 {\cal I}_{++} = \{\omega_A : \omega_A \in (\Delta, +\pi ]\}  = \hspace{1.3in} \\
\hspace{0.0in} = \left\{ 
\begin{array}{ccccccccccccccccccccc}
\{\omega_B : \omega_B & \in & (&0 &,&  -\Delta+\pi &]&\}, \hspace{0.1in} \mbox{if} \hspace{0.1in} \eta > 0  \\ 
\{\omega_B : \omega_B & \in & (&-\pi &,&  -\Delta&]&\}, \hspace{0.1in} \mbox{if} \hspace{0.1in} \eta \le 0  \\
\end{array}
\right.
\end{eqnarray*}

\begin{eqnarray*}
 {\cal I}_{+-} = \{\omega_A : \omega_A \in (0, \Delta ]\} = \hspace{1.3in} \\
\hspace{0.0in} = \left\{ 
\begin{array}{ccccccccccccccccccccc}
\{\omega_B : \omega_B & \in & (&-\Delta&,&  0 &]&\}, \hspace{0.1in} \mbox{if} \hspace{0.1in} \eta > 0  \\ 
\{\omega_B : \omega_B & \in & (&-\Delta+\pi &,&  \pi &]&\}, \hspace{0.1in} \mbox{if} \hspace{0.1in} \eta \le 0  \\
\end{array}
\right.
\end{eqnarray*}

\begin{eqnarray*}
{\cal I}_{--} = \{\omega_A : \omega_A \in (\Delta-\pi, 0 ]\}  = \hspace{1.0in} \\
\hspace{0.0in} = \left\{ 
\begin{array}{ccccccccccccccccccccc}
\{\omega_B : \omega_B & \in & (&-\pi&,&  -\Delta &]&\}, \hspace{0.1in} \mbox{if} \hspace{0.1in} \eta > 0  \\ 
\{\omega_B : \omega_B & \in & (&0 &,& -\Delta+\pi &]&\}, \hspace{0.1in} \mbox{if} \hspace{0.1in} \eta \le 0  \\
\end{array}
\right.
%{\cal I}_{-+} & = & \{\omega_A : \omega_A \in & (&-\pi,& \Delta-\pi &]&\} & = \\ 
%& = & \{\omega'_B : \omega'_B \in & (&-\Delta+\pi,& +\pi &]&\} &
\end{eqnarray*}

\begin{eqnarray*}
{\cal I}_{-+} = \{\omega_A : \omega_A \in (-\pi, \Delta-\pi ]\}  = \hspace{1.0in} \\
\hspace{0.0in} = \left\{ 
\begin{array}{ccccccccccccccccccccc}
\{\omega_B : \omega_B & \in & (&-\Delta+\pi&,& \pi &]&\}, \hspace{0.1in} \mbox{if} \hspace{0.1in} \eta > 0  \\ 
\{\omega_B : \omega_B & \in & (&-\Delta &,& 0 &]&\}, \hspace{0.1in} \mbox{if} \hspace{0.1in} \eta \le 0  \\
\end{array}
\right.
% & = \\ 
%& = &  &
\end{eqnarray*}
where we have assumed without any loss of generality that $0 \le \Delta \le \pi$.

In each one of these four segments, the two measurements are fully correlated or anti-correlated:

\begin{itemize}
\item If $\eta \in (0, +\pi]$, 

\begin{eqnarray}
\begin{array}{cccccccccccccccccc}
\left. S^{(A)}_{\Omega_{\alpha}}(\omega_A,\eta) \cdot S^{(B)}_{\Omega_{\beta}}(\omega_B,\eta)\right|_{{\cal I}_{++} \bigcup {\cal I}_{--}} & = &+1, \\ 
\left. S^{(A)}_{\Omega_{\alpha}}(\omega_A,\eta) \cdot S^{(B)}_{\Omega_{\beta}}(\omega_B,\eta)\right|_{{\cal I}_{+-} \bigcup {\cal I}_{-+}} & = &-1,
\end{array}
\end{eqnarray}

\item If $\eta \in (-\pi, 0]$,

\begin{eqnarray}
\begin{array}{cccccccccccccccccc}
\left. S^{(A)}_{\Omega_{\alpha}}(\omega_A,\eta) \cdot S^{(B)}_{\Omega_{\beta}}(\omega_B,\eta)\right|_{{\cal I}_{++} \bigcup {\cal I}_{--}} & = &-1, \\ 
\left. S^{(A)}_{\Omega_{\alpha}}(\omega_A,\eta) \cdot S^{(B)}_{\Omega_{\beta}}(\omega_B,\eta)\right|_{{\cal I}_{+-} \bigcup {\cal I}_{-+}} & = &+1,
\end{array}
\end{eqnarray}

\end{itemize}
\

It is straighforward to notice that 

\begin{equation}
\mu\left({\cal I}_{++} \bigcup {\cal I}_{--}\right)-\mu\left({\cal I}_{+-} \bigcup {\cal I}_{-+}\right) = \cos(\Delta),
\end{equation}
where $\mu(\cdot)$ denotes the normalized measure over the torus according to the probability density distribution (\ref{density_states}). Hence,

\begin{itemize}
\item  Over the subpopulation of states with $\eta > 0$,

\begin{equation}
\langle S^{(A)}_{\Omega_{\alpha}} \cdot S^{(B)}_{\Omega_{\beta}} \rangle = \ \ \ \cos(\Delta),
% = \ \ \cos(\Delta_{AB}+\Gamma)
\end{equation}

\item Over the subpopulation of states with $\eta \le 0$,

\begin{equation}
\langle S^{(A)}_{\Omega_{\alpha}} \cdot S^{(B)}_{\Omega_{\beta}} \rangle = -\cos(\Delta).
% = -\cos(\Delta_{AB}+\Gamma)
\end{equation}

\end{itemize}  
Therefore, over the whole population the two measurements are completely uncorrelated,

\begin{equation}
\langle S^{(A)}_{\Omega_{\alpha}} \cdot S^{(B)}_{\Omega_{\beta}} \rangle = 0, 
\end{equation}
since each one of the two populations $\eta > 0$ and $\eta \le 0$ happens with probability $1/2$. 

Furthermore, if we now define the outcome of the measurement apparatus of observer C as:

\begin{eqnarray}
S^{(C)}_{\Omega_{\gamma}}(\omega,\eta) = \left\{
\begin{array}{ccccc}
+1, \ \ \mbox{if} \ \ \eta > 0, \\
-1, \ \ \mbox{if} \ \ \eta \le 0, \\
\end{array}
\right.
\end{eqnarray}
we obtain that the three-particles correlation is given by:

\begin{equation}
\langle S^{(A)}_{\Omega_{\alpha}} \cdot S^{(B)}_{\Omega_{\beta}} \cdot S^{(C)}_{\Omega_{\gamma}} \rangle = \cos(\Delta) 
%= \cos(\Delta_{AB}+\Gamma),
\end{equation}
which reproduces the quantum mechanical predictions for the GHZ state.
\\

Indeed, by an appropriate choice of the two subpopulations over which the outcome of the measurement apparatus of observer C is defined either positive or negative we can arbitrarily modify the angle $\Phi$ in (\ref{the_law},\ref{the_frame}) and, hence, the actual value for the three-particles correlation. For the sake of concreteness we shall show now how we can choose two subpopulations such that over one of them $S^{(C)}_{\Omega^*_{\gamma}}=+1$ and over the other $S^{(C)}_{\Omega^*_{\gamma}}=-1$ and 

\begin{equation}
\label{the_proof}
\langle S^{(A)}_{\Omega_{\alpha}} \cdot S^{(B)}_{\Omega_{\beta}} \cdot S^{(C)}_{\Omega^*_{\gamma}} \rangle = +1.
\end{equation}  

We shall identify the two new subpopulations by new coordinates $\eta^* \le 0$ and $\eta^* > 0$, where

\begin{eqnarray}
\nonumber
\eta^* & = & \left\{
\begin{array}{cccccccccccccccccc}
- &\hspace{-0.07in}\eta,  & & \hspace{0.015in} \ \mbox{if} \hspace{0.05in} & (\omega, \eta) & \in & {\cal I}_{+-} & \bigcup & {\cal I}_{-+}, \\
&\hspace{-0.07in}\eta,  & & \hspace{0.015in} \ \mbox{if} \hspace{0.05in} & (\omega, \eta) & \in & {\cal I}_{++} & \bigcup & {\cal I}_{--}.
\end{array}
\right.
\end{eqnarray}
Simultaneously, we rename the angular coordinates of the hidden states as follows:

\begin{eqnarray}
\nonumber
\omega^*_A & = & \left\{
\begin{array}{cccccccccccccccccc}
\hspace{-0.02in}-\omega_A, & & & \hspace{0.05in} \ \mbox{if} \hspace{0.05in} & (\omega, \eta) & \in \ &  {\cal I}_{+-} & \bigcup & {\cal I}_{-+}, \\
\hspace{-0.05in}-\omega_A & \hspace{-0.075in} + & \hspace{-0.055in} \pi, & \hspace{0.05in} \ \mbox{if} \hspace{0.05in} & (\omega, \eta) & \in \ & {\cal I}_{++} & \bigcup & {\cal I}_{--}.
\end{array}
\right. \\
\nonumber
\omega^*_B & = & \left\{
\begin{array}{cccccccccccccccccc}
\hspace{0.04in} \omega_B & \hspace{-0.075in} + & \hspace{-0.055in} \pi,& \hspace{0.05in} \ \mbox{if} \hspace{0.05in} & (\omega, \eta) & \in \ &  {\cal I}_{+-} & \bigcup & {\cal I}_{-+}, \\
\hspace{0.09in} \omega_B, & & & \hspace{0.045in} \ \mbox{if} \hspace{0.05in} & (\omega, \eta) & \in \ & {\cal I}_{++} & \bigcup & {\cal I}_{--}.
\end{array}
\right. \
\end{eqnarray}
such that the definition of the outcomes of measurement apparatus A and B remains invariant for all hidden configurations:
\begin{equation}
S(\omega^*,\eta^*)=\mbox{sign}(\omega^*) \cdot \mbox{sign}(\eta^*)=\mbox{sign}(\omega) \cdot \mbox{sign}(\eta)=S(\omega,\eta).
\end{equation} 

Furthermore, it is straightforward to check that neither the probability to hapen of each one of these configurations gets modified, since the probability density (\ref{density_states}) remains invariant,

\begin{equation}
g(\omega) = g(-\omega) = g(\omega+\pi),
\end{equation}
as well as the actual actual value of the defined outcomes of strong measurement performed by observers A and B.
\\

\noindent It is now straightforward to check that the outcomes of the strong polarization measurements of particles A and B along directions $\Omega_{\alpha}$ and $\Omega_{\beta}$ are:

\begin{itemize}
\item   fully correlated $\langle S^{(A)}_{\Omega_{\alpha}} \cdot S^{(B)}_{\Omega_{\beta}} \rangle = +1$, 
over the subpopulation of states with $\eta^* > 0$.

\item fully anticorrelated $\langle S^{(A)}_{\Omega_{\alpha}} \cdot S^{(B)}_{\Omega_{\beta}} \rangle = -1$, over the subpopulation of states with $\eta^* \le 0$.

\end{itemize} 
Therefore, by identifying these two new subpopulations as corresponding to postive or negative outcomes for the measurement of the polarization of particle C along the direction $\Omega^*_{\gamma}$:

\begin{eqnarray}
S^{(C)}_{\Omega^*_{\gamma}}(\omega^*,\eta^*)=+1, \hspace{0.3in} & \mbox{if} \hspace{0.3in} \eta^* > 0, \\
S^{(C)}_{\Omega^*_{\gamma}}(\omega^*,\eta^*)=-1, \hspace{0.3in} & \mbox{if} \hspace{0.3in} \eta^* \le 0,
\end{eqnarray}
we obtain that the three-particles correlation is now given by eq. (\ref{the_proof}) as we had advanced.
%\begin{equation}
%\langle S^{(A)}_{\Omega_{\alpha}} \cdot S^{(B)}_{\Omega_{\beta}} \cdot S^{(C)}_{\Omega'_{\gamma}} \rangle = +1. 
%\end{equation}
%Our model thus reproduces the predictions of quantum mechanics for the GHZ polarization state of three entangled particles. 
\\

{\bf 6.} We have shown in this paper that the argument behind the renowned GHZ paradox crucially relies on an implicit assumption that is not required by fundamental physical principles. Thus the argument can be overcome if we give up this unnecessary assumption. Namely, the argument put forward by Greenbereger, Horne and Zeilinger thirty years ago \cite{GHZ} implicitly assumes that there exists an absolute angular frame of reference with respect to which we can define the polarization properties of the hypothetical hidden configurations of the entangled particles, as well as the orientation of the measurement apparatus that test them. We remarked in this paper that such an absolute frame of reference does not necessarily exist. Indeed, only the relative orientation of the set of measurement apparatus with respect to a reference setting is an actual physical degree of freedom, while the individual orientations of each one of the apparatus are spurious gauge degrees of freedom. Moreover, the polarization properties of each one of the particles can only be properly defined with respect to the orientation of the corresponding measurement apparatus. With these observations in mind we built an explicitly local and realistic model of hidden variables that complies with the 'free-will' assumption and reproduces the predictions of quantum mechanics for the GHZ state of three entangled particles. This model resembles closely the model of hidden variables for the Bell's polarization states of two entangled particles that we recently described in \cite{oaknin3}.

\end{document}